# Dusty Magnetohydrodynamics in Star Forming Regions.

S. VAN LOO[1], S.A.E.G. FALLE[2], T.W. HARTQUIST[1], O. HAVNES[3,4,5], and G.E. MORFILL[4]

[1] School of Physics and Astronomy, University of Leeds, Leeds LS2 9JT, United Kingdom

[2] Department of Applied Mathematical Sciences, University of Leeds, Leeds LS2 9JT, United Kingdom

[3] Department of Physics and Technology, University of Tromsö, 9037 Tromsö, Norway

[4] Max-Planck-Institut für extraterrestrische Physik, 85740 Garching, Germany

[5] University Center (UNIS) , 9071 Longyearbyen, Svalbard, Norway.

**Abstract.** Star formation occurs in dark molecular regions where the number density of hydrogen nuclei , $n_H$, exceeds $10^4$ cm$^{-3}$ and the fractional ionization is $10^{-7}$ or less. Dust grains with sizes ranging up to tenths of microns and perhaps down to tens of nanometers contain just under one percent of the mass. Recombination on grains is important for the removal of gas phase ions, which are produced by cosmic rays penetrating the dark regions. Collisions of neutrals with charged grains contribute significantly to the coupling of the magnetic field to the neutral gas. Consequently, the dynamics of the grains must be included in the magnetohydrodynamic models of large scale collapse, the evolution of waves and the structures of shocks important in star formation.

# 1. Introduction

Trumpler [1] in 1930 and Stebbins, Huffer and Whitford [2,3] in 1934 and 1939 reported the results of photometric studies that demonstrated that dark "holes" in the Galaxy, noted by Herschel close to 150 years before, are due to obscuration by interstellar dust. In the first part of his important May 1941 paper on the charge carried by interstellar dust, Spitzer [4] wrote "The presence of dust particles implies that the physical state of such a medium may be somewhat different than that discussed in the classical work by Eddington, in which only atoms were assumed to be abundant". Despite Spitzer's early insight, the first papers on the effects of dust on the magnetohydrodynamic behavior of interstellar matter appeared only about three decades ago.

In 1997 Hartquist, Pilipp and Havnes [5] published an introductory review of work on dusty plasma in interstellar clouds and star forming regions. While numerous advances have since occurred, we refer the reader to that article for a more detailed presentation of many of the basic physical processes and equations important for the area. Though we mention later work here, as well as give a qualitative explanation of some of the key concepts, we have not attempted to be comprehensive. Our current interest in multi-fluid shocks in star forming regions has guided our choice of emphasis, but we do mention some other progress.

Section 2 contains a brief description of the processes establishing the fractional ionization in dark regions in dense cores, molecular structures that are progenitors of stars; recombination on grains is significant. Section 3 provides a description of the role that dust plays in ambipolar diffusion, the motion of ions relative to neutrals resulting from magnetic field gradients, in dense cores and protostellar formation. Section 4 contains a brief summary of the results of an early study of the effects of dust on the damping of Alfvén waves in dense cores. Section 5 is a selective short review of work on the effects of dust on the structure of shocks assumed to propagate perpendicularly to the magnetic field in star forming regions. Section 6 contains a similar review of work published prior to 2009 on shocks propagating obliquely to the upstream magnetic field. Section 7 concerns more recent work on such shocks. Section 8 concludes the paper.

**2. The Fractional Ionization in Dark Star Forming Regions.**

Stars form in dense cores, which are molecular condensations embedded in more diffuse molecular gas that is many tens or more times less dense than the dense cores. Most, if not all, of the more diffuse gas surrounding dense cores is in translucent clumps making up Giant Molecular Clouds (GMCs) or GMC complexes (see, e.g., section 2 of [5] for a review). An $n_H$ (i.e. the number density of hydrogen nuclei) range of $10^4 - 10^5$ cm$^{-3}$ is typical of dense cores in which solar-like stars form, but an $n_H$ range of $10^6 - 10^7$ cm$^{-3}$ is typical of regions in which high-mass stars, those that contain at least 8 solar masses, are born [6]. Of course, star formation leads to values of $n_H$ greatly exceeding these, but most of this paper concerns phenomena occurring in the $n_H$ range of $10^4 - 10^7$ cm$^{-3}$.

Oppenheimer and Dalgarno [7], Gail and Sedlmayr [8], Elmegreen [9], Draine and Sutin [10] and Umebayashi and Nakano [11] are amongst those who have contributed to the development of our understanding of the ionization structure and the calculation of the grain charge in dark star forming regions. Many of the considerations have much in common with those addressed by Shukla and Mamun[12].

Cosmic ray ionization of $H_2$ at a rate of the order of $10^{-17} - 10^{-16}$ s$^{-1}$ leads to $H_2^+$, which reacts with $H_2$ to form $H_3^+$. $H_3^+$ in turn reacts with neutral species including O, $H_2O$ and CO (the most abundant gas phase molecule other than $H_2$ in most circumstances) to form molecular ions. Dissociative recombination is a major mechanism for removing molecular ions in dense cores, but its effectiveness is moderated if metals, like magnesium and sodium, are not too depleted onto grains. Charge transfer between molecular ions and such metals produces metallic ions. The gas phase process removing them is radiative recombination, which is many orders of magnitude slower than dissociative recombination. The primary mechanism removing metallic ions is recombination on grains.

At the typical temperatures of dense cores of the order of 10 K and typical core densities, most grains will carry one negative charge if nearly all of the grain material is in grains of sizes of around 0.1 μm. Considerable uncertainty concerning the grain size distribution exists. Certainly in more diffuse clouds many grains of much smaller size exist. However, in dense, dark regions the formation of ice mantles by the depletion of gas phase species may lead to almost all grains having sizes of about 0.1 μm.

The grain charge probability distribution function and the fractional ionization must be calculated self-consistently. Some of the authors cited above and many others have done so. Fig.7 of [5] shows results for the fractional abundances of electrons, gas phase ions, grains carrying a single negative charge, grains carrying a single positive charge and neutral grains as functions of $n_H$ ranging from $10^4$ to $10^{14}$ cm$^{-3}$. All grains were assumed to be spherical with radii of 0.1 μm and to have a fractional abundance relative to $n_H$ of $4\times10^{-12}$. The cosmic ray induced ionization rate and the temperature were taken to be $10^{-17}$ s$^{-1}$ and 10 K, respectively.

For $n_H$ up to about $10^9$ cm$^{-3}$, the fractional ionization drops roughly as $n_H^{-\frac{1}{2}}$ from $3\times10^{-8}$. The fractional abundances of ions and electrons are nearly equal and almost all grains are negatively charged. At $n_H$ above $10^{10}$ cm$^{-3}$ and below $10^{12}$ cm-3 most grains are neutral and the fractional abundances of ions and of negatively charged grains are nearly equal, while the fractional abundance of electrons is lower. At $n_H$ above $10^{12}$ cm$^{-3}$, the fractional abundances of negatively charged and positively charged grains are nearly equal, while the fractional abundance of ions is lower.

**3. Ambipolar Diffusion in Dense Core Collapse.**

There has been considerable debate about the relative roles, in dense core formation and evolution, of magnetically regulated, gravitationally driven collapse and non-linear magnetohydrodynamic processes that occur even in the absence of gravity [13]. Ambipolar diffusion is important for the relevant non-linear magnetohydrodynamic processes [14] and grains must be included in the treatment of its effect on the those processes. However, the bulk of the studies on the importance of grains for ambipolar diffusion in star formation have been performed for a picture in which a dense core is supported by a combination of thermal pressure and magnetic forces. In this picture collapse occurs as magnetic force drives the charged particles to move relative to the neutrals, thus reducing the magnetic contribution to the support.

The timescale for the decrease in the magnetic support depends upon the force per unit volume due to collisions between neutrals and charged species. Baker [15], Elmegreen [9], and Nakano and Umebayashi [16] were the first to recognize the importance of the contribution of collisions between neutrals and charged grains to this force per unit volume and to calculate its effect on the ion slip velocity and ambipolar diffusion/collapse time scale.

We summarize the conditions [5] under which grain-neutral collisions significantly affect the velocity of electrons relative to the neutrals in a region undergoing gravitational induced collapse as a consequence of ambipolar diffusion. We assume that the charges on all grains are equal and do not fluctuate. $\Omega$ is the magnitude of the grain gyrofrequency, $v_{jk}$ is the frequency at which a single particle of species $j$ undergoes collisions with particles of species $k$. The subscripts can be $n$, $g$ or $i$ indicating neutrals, grains and ions, respectively. If $v_{ng} \ll \Omega$, the condition is

$$v_{ng} \gg v_{ni}$$

If $v_{ng} \gg \Omega$, the condition is

$$\frac{v_{ni}^2}{v_{ng}^2} \frac{v_{gn}^2}{\Omega^2} \ll 1$$

In a series of papers Ciolek and Mouschovias, e.g. [17], and Tassis and Mouschovias, e.g.[18], have presented the results of the most thorough studies of ambipolar-diffusion regulated, gravitationally driven collapse. They have considered the collapse of thin disks. They performed self-consistent calculations of the fractional ionization and probability distribution function for the charges on grains. Though all grains were taken to have the same size, fluid equations were included for grains of each charge. For the low temperatures involved only three grain charges needed to be considered.

**4. The Effects of Dust on Alfvén Waves.**

Of course, the most famous paper on the effects of dust on waves is that of Rao, Shukla and Yu [19] on dust acoustic waves. By now the literature on waves in dusty plasmas is immense. Here, we consider results of an early study [20] of the propagation of Alfvén waves in dusty star forming regions in order to draw attention to some of the relevant physics. Waves are important as they contribute to the widths of the observed line profiles in star forming regions and their non-linear evolution may contribute to the production of dense cores. Kulsrud and Pearce [21] considered the damping of Alfvén waves in a weakly ionized dustless plasma due to ion-neutral friction. An analysis restricted to linear waves propagating parallel to the large scale magnetic field leads to an approximate dispersion relation of

$$k^2 = \frac{4\pi\rho_i}{B_0^2}\omega^2\left(1 + \frac{v_{in}}{v_{ni} - i\omega}\right)$$

where $k$ is the complex wavenumber, $\omega$ is the real angular frequency, $B_0$ is the strength of the large-scale magnetic field and $\rho_i$ is the mass density of ions.

As grain-neutral collisions in star forming regions had been recognized to be important for ambipolar diffusion, the initiation of studies of their role in wave propagation was a natural step.

One set of results from a linear analysis [20] of Alfvén waves propagating parallel to the large-scale magnetic field in a dense core were for $n_H = 2 \times 10^4$ cm$^{-3}$, $T=20$ K and $B_0=10^{-4}$ G. The dust grains were assumed to contain one percent of the mass and have uniform radii of 0.1 μm and mass density 1 g cm$^{-3}$. The number densities of ions and electrons were taken to be $10^{-3}$ cm$^{-3}$. Results for the interesting range of $\omega$ of $10^{-4}$ to $10^{-5}$ y$^{-1}$ show a reduction in the damping rate of a factor of 2 or 3 for cases in which the dynamics of neutral and negatively charged grains are included compared to those in which they are excluded. The reduction would have been much larger for different reasonable assumptions for the ion and electron number densities.

Van Loo et al. [14] have found that the nonlinear development of waves with $\omega$ in about this range is sensitive to the degree of collisional coupling between charged and neutral fluids. In the near future, the inclusion of at least two grain fluids in treatments of the nonlinear development of waves in star forming regions will be expected.

The effect of the charge fluctuations of grains on the dissipation of the waves was revealed in the initial study of Alfvén waves in dusty media [20]. The influence of charge fluctuations on wave damping has been studied subsequently in many contexts.

## 5. Perpendicular shocks.

Once stars form, their outflows affect the environment. The interaction of the outflows with ambient molecular gas leads to the formation of shocks that may lead to the compression of inhomogeneities to trigger more star formation or possibly to the disruption of them resulting in the termination of stellar birth. The shocks also alter the chemical and dust contents of the star forming regions, in part by sputtering dust grains.

Two key papers on the nature of shocks in star forming regions are those by Draine [22] and Draine, Roberge and Dalgarno [23].

Shocks propagating perpendicular to the large-scale magnetic field are fast-mode shocks. We will assume that the magnetic pressure is significantly greater in the preshock medium than the thermal pressure. Thus, the fast-mode and Alfvén speeds are about equal. Two upstream Alfvén speeds are relevant. One is the Alfvén speed of low frequency waves, $V_{Al}$, which is determined by the upstream magnetic field strength, $B_0$, and the sum of the mass densities of all species. The other is the Alfvén speed of high frequency waves, $V_{Ah}$, which depends on $B_0$ and the mass density of charged species that are well-coupled to the magnetic field. If $V_S$, the shock speed, exceeds $V_{Ah}$ the shock will have jumps in all fluid parameters. However, for a range of $V_S < V_{Ah}$ but $> V_{Al}$, the flow variables in all fluids are continuous. We consider only such C-type shocks. In a frame comoving with the C-type shock, the charged particles decelerate from the shock velocity further upstream than the neutrals do. This creates a precursor. Requiring the gradient in the magnetic pressure to be comparable to the drag force per unit volume on ions due to collisions with neutrals, one can find that the precursor thickness is roughly

$$\Delta = \frac{V_{Al}^2}{2\nu_{ni} V_S}$$

$\Delta$ is the lengthscale over which dissipation occurs, and the smaller $\Delta$ is the hotter the gas becomes. If grains were present and well-coupled to the magnetic field, the expression for $\Delta$ would contain the sum of $\nu_{ni}$ and $\nu_{ng}$, rather than $\nu_{ni}$ alone. However, collisions tend to decouple the grains from the magnetic field, an effect considered by Draine [22] and Draine, Roberge and Dalgarno [23] whose models of perpendicular shocks are reliable for cases in which the preshock value of $n_H$ is $\leq 10^6$ cm$^{-3}$.

At higher densities a self consistent calculation of the average charge on the grains and the fractional ionization is required to obtain accurate results for $v_{ni}$ and for the effect that grains have on the shock structure. Pilipp, Hartquist and Havnes [24] included such a calculation in each of their models of perpendicular shocks. They also included fluid equations for the grains rather than adopt a simpler approximation to calculate the effects of dust [22, 23]. This led them to discover a run-away process operating in perpendicular shocks for which the preshock value of $n_H$ is $10^7$ cm$^{-3}$ or higher. At such densities, the ratio of $n(e)/|Z_g|n_g$ drops below unity within the precursors of sufficiently fast shocks. $n(e)$ is the electron number density, $|Z_g|e$ is the magnitude of the average charge carried by grains and $n_g$ is the number density of grains. Once this ratio drops below unity, $|Z_g|$ begins to drop as there are insufficient electrons to continue charging the grains. Assume that the shock propagates in the x-direction and the magnetic field is in the y-direction and that $\Omega/v_{gn}$, the grain Hall parameter, is small. Then the grains separate from the other charged particles sufficiently to generate an x-component of the electric field with a magnitude given approximately by

$$E_x = \frac{m_g v_{gn} |v_{nx} - v_{gx}|}{|Z_g|e}$$

Here $m_g$ is the mass of a grain. This creates an ion drift velocity component in the z-direction with a magnitude of $cE_x/B_y$. As $|Z_g|$ drops, this component of the drift velocity increases. This causes an increase in the rate at which a grain experiences collisions with ions, which leads to a further drop in $|Z_g|$. Hence, a runaway occurs.

The recent work of Guillet, Jones and Pineau des Forêts [25, 26] represents a significant development in the modeling of perpendicular shocks in dusty star forming regions. In their work on C-type shocks [25], they adopted a hybrid approach. The gaseous species were described as fluids, whereas the trajectories and charges of many individual dust particles were followed. The treatment gave self-consistent results for the grain charges, gas phase ion and electron abundances and dynamics. Though the fluid approach of Pilipp et al. [24] is valid for a wide range of parameter space, the Guillet et al. [25] method is required in cases in which dust gyroradii are comparable to or larger than the scales on which variations of parameters vary in shocks.

**6. Steady-State Models of Oblique Shocks.**

Pilipp and Hartquist [27] adopted a fluid description of grain dynamics in studies of steady shocks propagating obliquely to the upstream magnetic field in dusty star forming regions. We will assume that a shock propagates in the x-direction and that the upstream magnetic field has x and y components but its z component is zero. They found that grain-neutral collisions lead to a rotation of the magnetic field in a C-type shock precursor around the x-direction.

The following considerations show why such rotation occurs. In the shock frame the z-component of the electric field, $E_z$, is $V_S B_{y0}/c$, where $B_{y0}$ is the y-component of the upstream magnetic field. Thus, there is a component of $\boldsymbol{E} \mathrm{x} \boldsymbol{B}$ drift in the x-direction. In the absence of collisions, the x-components of the $\boldsymbol{E} \mathrm{x} \boldsymbol{B}$ drift velocities of all species are equal. However, grain-neutral collisions are significant leading to a non-zero x-component of the Hall current. For a steady shock, the equation of charge conservation requires that the x-component of the total current is zero. Thus, a current parallel to the magnetic field having a x-component that cancels the Hall current component in the x direction must exist. Use of Ampere's Law shows that the current along the magnetic field generates a component of the magnetic field in the z direction.

By integrating from an upstream point in the downstream direction, Pilipp and Hartquist [27] succeeded in finding only intermediate-mode shock solutions. Such solutions are inadmissible [28].

Wardle [29] showed that integration in the downstream direction will not yield steady fast-mode solutions because the downstream state corresponds to a saddle point. He found fast-mode solutions by integrating upstream from the downstream state.

Chapman and Wardle [30] have extended this work and shown that the inclusion of PAHs leads to a drop in the gas phase electron abundance and enhanced rotation of the magnetic field. PAHs are nano-particles thought to be abundant in clouds more diffuse than star forming regions. As mentioned earlier, in such regions desorption of material from the gas phase may result in all particles growing to sizes close to 0.1 μm.

Integration in the upstream direction is not appropriate if conditions anywhere in a shock deviate from equilibrium. After shocked gas has cooled, the abundance of $H_2O$, an important coolant in shocked dense core material, remains far from its equilibrium value for many times the flow time through a shock. Other chemical species also have abundances that are far from their equilibrium values for considerable periods. Consequently, the calculation of shock structures by integration in the upstream direction is inappropriate, and the use of a time-dependent, rather than a steady-state, approach is necessary to overcome the difficulties found by Pilipp and Hartquist [27] and explained by Wardle [29]. Falle [31] has developed an appropriate time-dependent approach.

## 7. Time-Dependent Models of Oblique Shocks in Dusty Regions.

Van Loo et al. [32] have used the technique developed by Falle [31] to model oblique shocks in uniform density media. Figures 1 and 2 show results obtained for a shock that has evolved to a steady-state structure. As shown by Wardle [29] when field rotation is significant the trajectory in $B_y$, $B_z$ phase space corresponds to a spiral node in the vicinity of the upstream fast-mode state. As seen from Figures 1 and 2, the velocity structure is complicated where magnetic field rotation is significant.

The availability of a time-dependent code opens the possibility of studying shocks in inhomogeneous media. Star forming regions contain density structures on a variety of scales and, as mentioned in Section 4, the responses of such structures to shocks determine whether star formation induces further stellar birth or causes it to cease. Ashmore et al [33] used the Falle-Van Loo et al. code to study oblique shocks in inhomogeneous star forming regions. Results similar to those that they presented are displayed in Figure 3.

**8. Conclusion**

The major challenge in star formation theory will be the incorporation of the effects of dust in multidimensional, time-dependent magnetohydrodynamic simulations. The work reported in many of the papers cited in this brief review demonstrates the existence of a community that appreciates the role that dust plays in the phenomena investigated in simulations of the dynamics of star formation. However, so far the assumption of non-ideal magnetohydrodynamics has been relaxed in only a handful of the multidimensional numerical studies, e.g. [14]. Hard but interesting work remains.

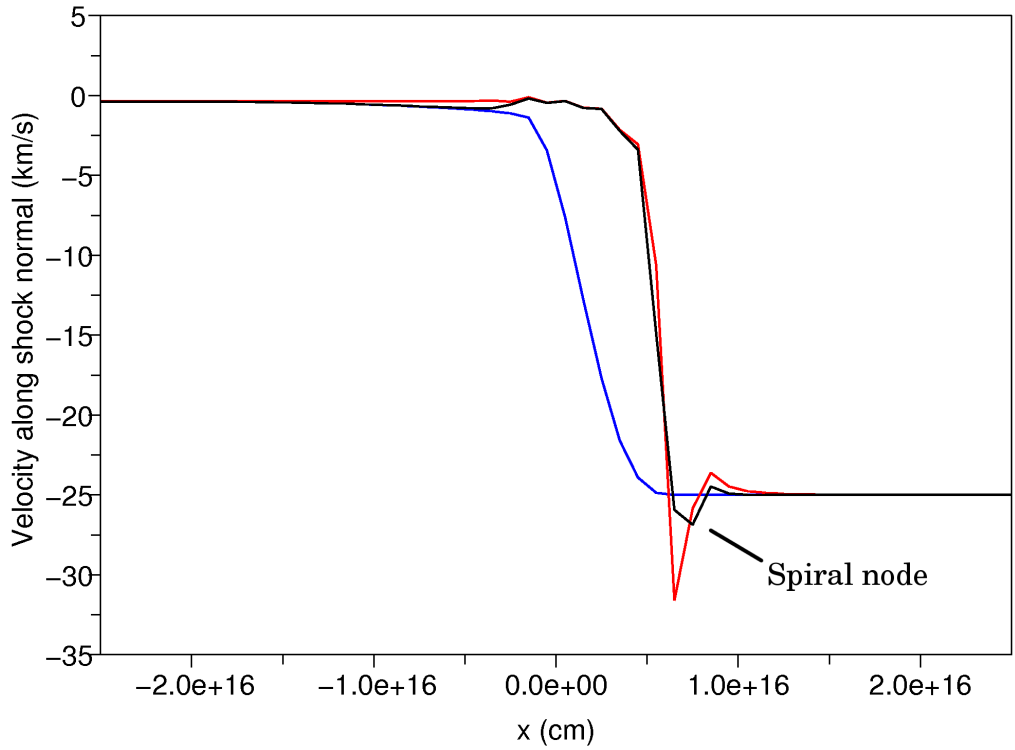

Figure 1. The velocity structure, in the shock frame, along the shock normal for a steady-state C-type shock propagating at 25 km/s through a homogeneous medium with $n_H = 10^5$ cm$^{-3}$. Each grain has a radius of 0.4 μm and a mass of 8x10$^{-13}$ g and one percent of the mass is in grains. The upstream magnetic field strength is 10$^{-4}$ G. The shock velocity is at an angle of 45º with respect to the upstream magnetic field. The blue line represents the neutral fluid, the red line the ion and electron fluids and the black line the grain fluid.

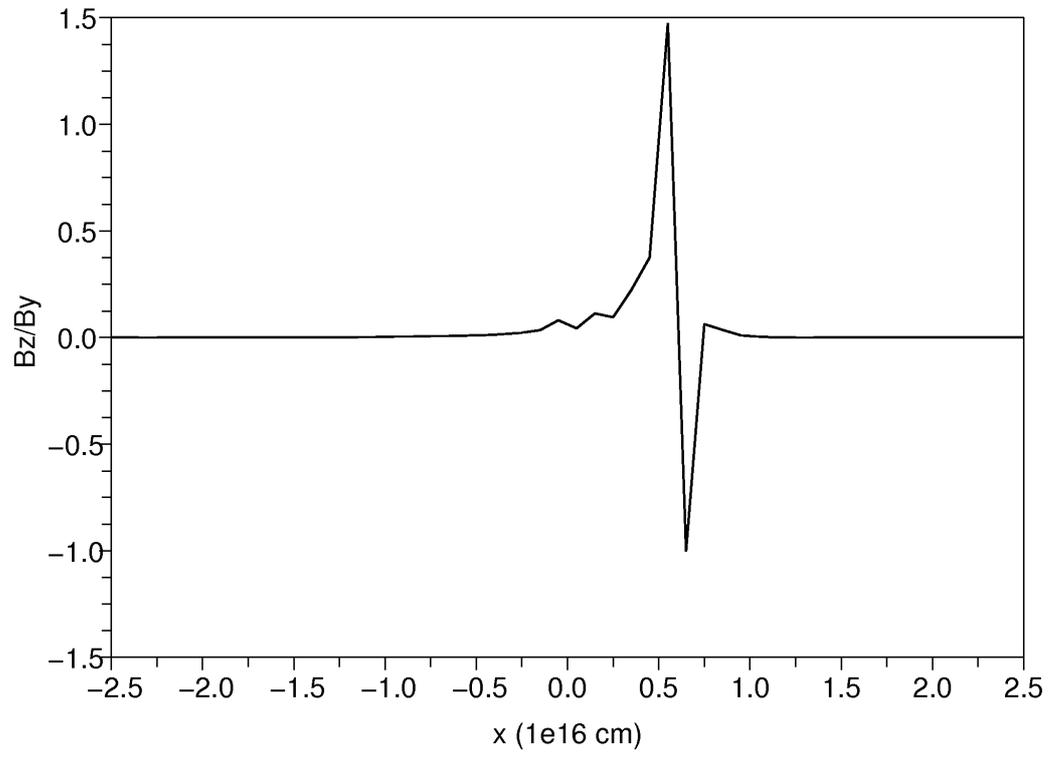

Figure 2. The ratio of $B_z$ to $B_y$ in the shock for which results are shown in Fig.1.

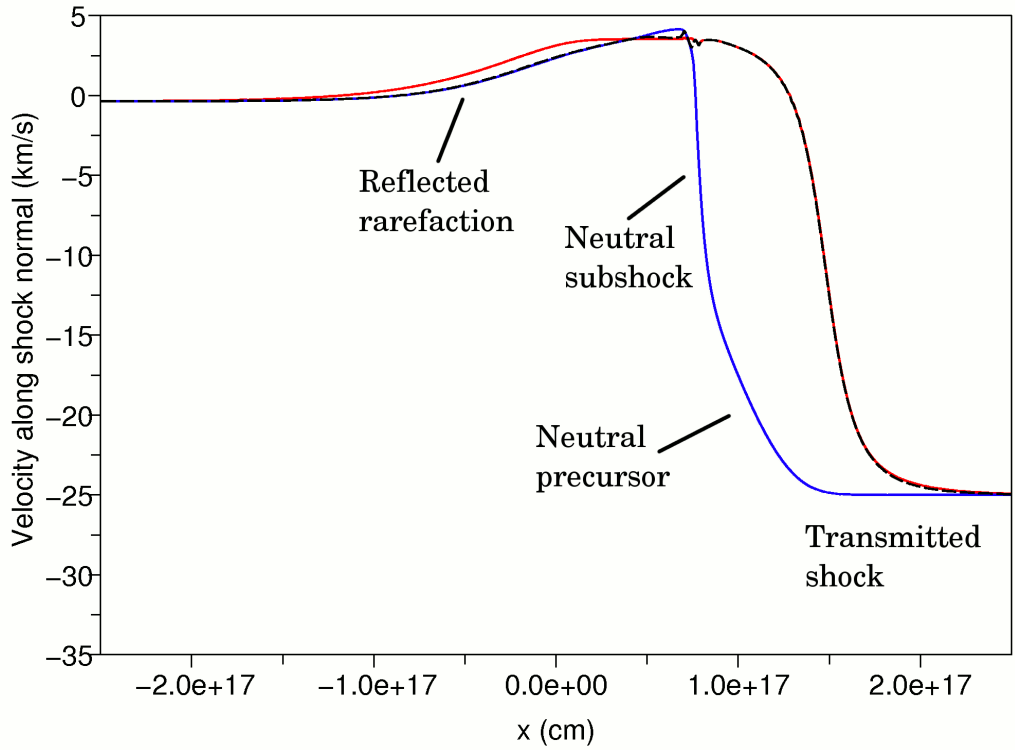

Figure 3. The velocity structure, in the initial shock frame, along the shock normal 5000 yr after the steady shock for which results are shown in Fig.1 propagates into a region of lower density (from $n_H = 10^5$ cm$^{-3}$ to $10^4$ cm$^{-3}$). The blue line represents the neutral fluid, the red line the ion and electron fluids and the black line the grain fluid.